\documentclass[aip,pof,preprint,superscriptaddress]{revtex4-2}
\usepackage{graphicx}
\usepackage{dcolumn}
\usepackage{bm}
\usepackage{xcolor}

\newcommand{\rev}[1]{#1}

\begin{document}


\title{\textbf{Data-driven surrogate models for forecasting experimentally measured fluid flows} 
}%

\author{Peter I Renn*}
\affiliation{
Division of Engineering and Applied Science \\
California Institute of Technology \\ Pasadena, CA, USA}%
\author{Emily H Palmer}
\affiliation{%
Albert Nerken School of Engineering\\
The Cooper Union for the Advancement of Science and Art \\
New York, NY, USA
}%
\author{Cong Wang}
\affiliation{%
College of Engineering \\ University of Iowa \\ Iowa City, IA, USA}%
\author{Morteza Gharib}
\affiliation{
Division of Engineering and Applied Science \\
California Institute of Technology \\ Pasadena, CA, USA}%
\date{10 October 2026 \\ **** Corresponding Author: \texttt{prenn@caltech.edu}}


\begin{abstract}
Data-driven modeling shows significant promise for faster-than-real-time forecasting of fluid flows. For real-world engineering applications (e.g., flow control), models must contend with limited, imperfect, and incomplete experimental measurements. \rev{However, most data-driven surrogates have been developed using and demonstrated entirely within the computational domain.} In this work, we present an analysis of data-driven surrogate models trained to forecast the time-evolution of experimentally measured cylinder wakes. Using a dataset of two-dimensional, two-component particle image velocimetry measurements, we train fully convolutional neural networks, U-Nets, Fourier neural operators, and dynamic mode decomposition-based models to forecast the development of experimentally measured velocity fields. To characterize data-driven approaches contending with transient flow features and limited, imperfect observations, the development of predictions over extended forecast horizons is examined at a fixed Reynolds number ($Re=590$). Next, models are trained at a range of Reynolds numbers ($Re =$ 230 to $Re =$ 2920) to investigate the impact of increasingly turbulent and three-dimensional flow phenomena on forecast quality. \rev{Through a triple decomposition of measured and forecast velocity fields, we find that error is dominated by the fluctuating velocity component, with all surrogates exhibiting a decay in estimated planar turbulent kinetic energy. The characteristics of this decay are architecture-dependent. Regardless, we find that the experimentally trained surrogate models can provide meaningful predictions over short time horizons, propagate low-frequency dynamics over longer forecast periods, and achieve faster-than-real-time evaluation in the considered contexts. These results highlight the opportunities and challenges that remain for data-driven fluid dynamics in real-world, real-time engineering applications.}

\end{abstract}

\maketitle

\section{Introduction}
Data-driven models have the potential to change the way engineered systems interact with non-linear fluid dynamics by enabling real-time reconstruction, prediction, and control of fluid flow \citep{vinuesa2023transformative, brunton2020}. While not suitable as replacements for physics-based numerical solvers commonly used in computational fluid dynamics, the evaluation speed of data-driven surrogate models could facilitate meaningful short-term predictive forecasts of fluid flows in faster-than-real-time. Such models could help realize predictive flow-control capabilities in many real-world engineering applications.

A wide array of data-driven methods for predicting the time evolution of fluid flows have been proposed. For example, \citet{srinivasan2019} and \citet{nakamura2021} applied neural network approaches to computational simulations of turbulent shear layers and three-dimensional channel flows, respectively. In both cases, the data-driven models match statistical quantities of the flow well but struggle to make accurate instantaneous predictions. \citet{wu2022} proposed a multi-resolution convolutional interaction network to make temporal predictions for cylinder flow data from eddy-viscosity turbulence models in the moderate subcritical regime, but the model was outperformed by basic linear dynamic mode decomposition for a cylinder flow with constant inlet velocity. \citet{fukami2021} combined a convolutional neural network auto-encoder with sparse identification of nonlinear dynamics (i.e., SINDy) to model laminar cylinder flow data generated by direct numerical simulation, as well as a shear flow model. This method was capable of learning the latent dynamics of both cases, however, the resulting models are sensitive to noisy observations and require problem-specific handling of learning parameters.  \citet{wang2020TF} proposed a deep learning framework inspired by the hybrid RANS-LES coupling, where velocity fields are decomposed into three different scales to improve prediction accuracy.

Particular interest has been shown in predicting fluid flows through operator learning frameworks, such as Fourier neural operators (FNOs) \citep{li2021fno}. \citet{wen2022ufno} successfully applied an FNO variant for predicting pressure buildup and gas saturation in multiphase flows. \citet{li2023ufno_turbulence} proposed another FNO variant for predicting isotropic turbulence, finding long-term predictions to be stable. \citet{han2025hybrid} developed a hybrid neural operator network to predict unsteady fluid flows, outperforming vanilla FNO by integrating convolutional long short-term memory and Fourier layers into a single model. 

Despite this interest, previous approaches have been largely limited to training on and predicting numerically-generated data. While some works have addressed the impact of artificially injected noise, \rev{this does not adequately represent or characterize the limitations inherent to real-world observations of fluid flows. Measurements} can often provide only in-plane dynamics (e.g., two-dimensional information in a three-dimensional flow) with limited resolution in both time and space, inherently failing to provide the full state information. 

In this paper, we examine data-driven methods as real-time capable surrogate models for the prediction of experimentally measured turbulent cylinder wakes in the context of realistic but imperfect physical observations. We seek to elucidate the relationship between data-driven model predictions, flow physics, and the limitations of observability in an experimental setting. \rev{Specifically, we demonstrate how  observability constraints (e.g., out-of-plane velocities) compound with limitations inherent to data-driven models. This reveals an architecture-dependent decay of forecast in-plane turbulent kinetic energy.} Using two-dimensional two-component particle image velocimetry (2D2C-PIV) data, common data-driven models are trained to recursively forecast the time evolution of turbulent cylinder wakes. Given the inherently three-dimensional nature of these flow fields, we explore how data-driven surrogates can learn the turbulent wake dynamics given incomplete and noisy state estimates.

First we explore the characteristics of model forecasts over an extended prediction horizon, equivalent to a full vortex shedding cycle, with a fixed Reynolds ($Re$) number of $Re=590$. Here we explore how the prediction quality varies over the trajectory of the forecast, including considerations of conservation of planar turbulent kinetic energy derived from the fluctuating component of the velocity fields. Then, the characteristics of the learned dynamics are analyzed at a range of Reynolds numbers ($Re=230$ to $Re=2920$) in the wake-transition and subcritical vortex shedding regime, where turbulent energy content and three-dimensional effects are increasing \citep{williamson1996review}. Here we specifically examine how data-driven models contend with the challenges associated with incomplete and imperfect observations in this regime, as well as the flow physics associated with high accumulated error. Through this analysis, we include several popular paradigms in data-driven modeling, including powerful linear baselines, to explore how various surrogate architectures can characterize the learned dynamics. We specifically consider popular data-driven architectures such as fully convolutional neural networks (FCNs), U-Nets, FNOs, and dynamic mode decomposition (DMD) based models. While this study is not intended to benchmark optimized model architectures, differences between various surrogates provide critical insight into the intrinsic limitations of modeling experimental fluid dynamics given realistic, incomplete observations. 

\rev{Data-driven models are found capable of forecasting large-scale flow dynamics over short time horizons with faster-than-real-time evaluation speeds. However, when faced with real-world fluid flow data, the fine-scale and transient dynamics in data-driven forecasts universally collapse over longer horizons, with the specific failure mode being architecture-dependent. This collapse results in forecast error saturating with respect to Reynolds number above $Re \approx 830$, despite increasingly turbulent dynamics. These findings indicate that known model and architecture limitations can be exacerbated by partially observable state information; future real-world fluid flow forecasting applications must carefully contend with these challenges.}


\section{Experimental Methods}

\subsection{Learning problem: cylinder wakes}

As one of the most well known and visually striking unsteady flows, vortex shedding in the wake of a cylinder has often been used to demonstrate data-driven methods in fluid mechanics \citep{hasegawa2020a,hasegawa2020b,morimoto2021}. The vortex shedding process has been studied in detail for decades, with several comprehensive review papers on the subject \citep{williamson1996review}. The phenomenon is first observed in the wake of a cylinder at a Reynolds number of approximately 50, at which point an instability occurs in the previously steady recirculation region of the wake. This instability results in a famously beautiful and well-ordered pattern of laminar alternating vortices convecting downstream away from the cylinder, known as a von Kármán vortex street. There has been some variation observed in defining the start of the transition from laminar to turbulent wakes, but it is generally placed around $Re$ = 190, at which point small-scale three-dimensional instabilities form and the transition to turbulence begins \citep{williamson1996threed}. The behavior of cylinder wakes from $Re$ = 300 to $Re$  = 200 000 was termed the ``irregular range" by \citet{roshko1958bluffbody} (otherwise known as the sub-critical regime). This regime is characterized by increasing three-dimensional effects, irregular velocity fluctuations, and the transition of the outer shear layer \citep{williamson1996review}. 

\begin{figure}
  \centerline{\includegraphics[width=\textwidth]{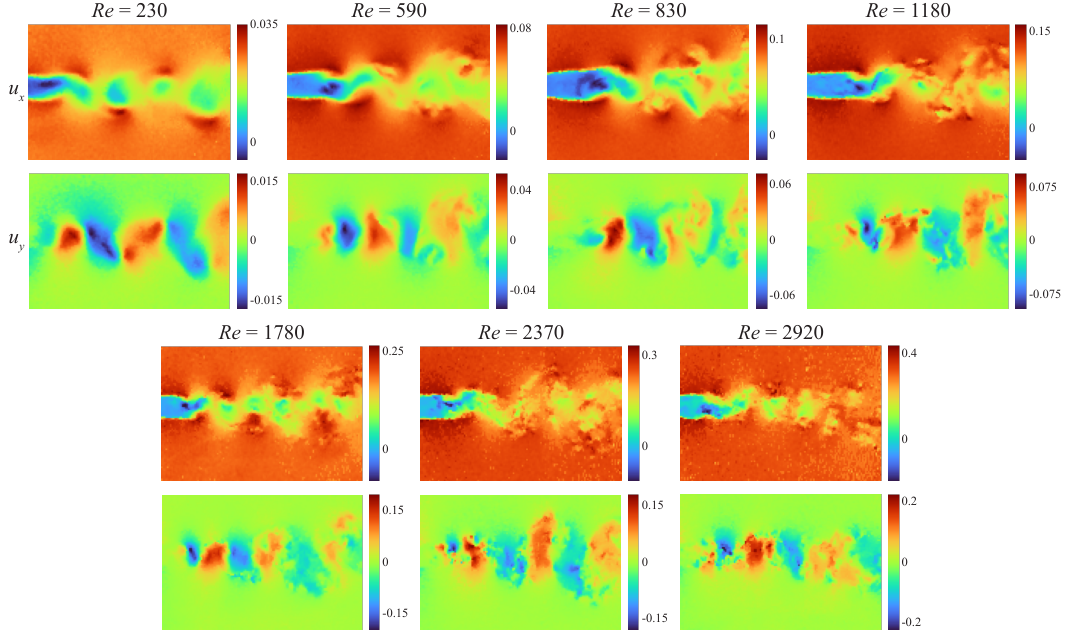}}
  \caption{Snapshots of velocity fields across the Reynolds numbers tested. Colorbar scales given in m/s.}
\label{fig:vel_ex}
\end{figure}

The experiments in this work were performed across a range of Reynolds numbers, from $Re$ = 230 to $Re$ = 2920, which can be seen in figure \ref {fig:vel_ex}. As expected, the flow appears increasingly turbulent and irregular with increasing Reynolds number. Speckling can be observed in the velocity fields measured at higher Reynolds numbers, which is attributable to an increase in high-frequency three-dimensional energy content that the measurements are unable to resolve. This setting enables evaluation of model performance across fluid phenomena that is increasingly challenging to both model and observe, while maintaining a common dominant flow feature in the primary vortex shedding periodicity.

\subsubsection{Data acquisition}
Data was collected using a free-surface water tunnel with test section dimensions of 0.15 m (W) $\times$ 0.15 m (H) $\times$ 0.61 m (L). An acrylic cylinder of diameter $D$ = 9.53 $\times$ 10$^{-3}$ m was mounted to span the walls of the test section approximately equidistant from the free surface and the tunnel floor. The cylinder aspect ratio was approximately 15.7.  Tests were performed at flow speeds ranging from $U$ = 0.024 m s$^{-1}$ to $U$ = 0.307 m s$^{-1}$.

\begin{figure}
  \centerline{\includegraphics[width=\textwidth]{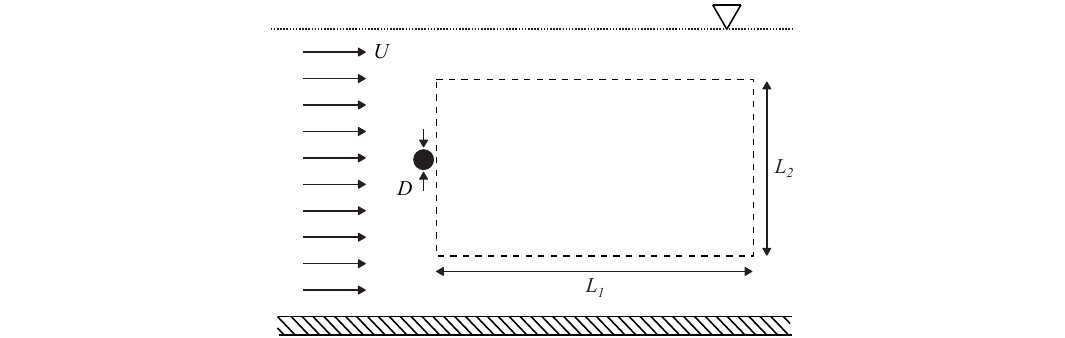}}
  \caption{Schematic of the imaging region of interest, outlined by the black dashed line, relative to the water tunnel and cylinder.}
\label{fig:exp_setup}
\end{figure}

Figure \ref{fig:exp_setup} depicts the geometry of the recorded region of interest relative to the tunnel boundaries and cylinder. The streamwise length of this region ($L_1$) was 0.125 m ($\approx 13D$) with transverse height ($L2$) of 0.084 m ($\approx 9D$). The flow in this region was illuminated by a continuous laser sheet approximately centered between the walls of the tunnel. High-speed video was recorded via an IDT XSM-3520 high-speed camera, with resolution 2144 px $\times$ 1440 px. To maintain a near-constant non-dimensional inter-sample time at various tunnel speeds, the frame rate of the camera was adjusted between Reynolds numbers. The non-dimensional formation time, given as $t^*=Ut/D$, was used to normalize the flow development between subsequent velocity fields across Reynolds numbers \citep{gharib2004}. Commercial software (PIVview by PIVTEC) was used to calculate velocity fields from  image pairs. During processing, the interrogation window was set to $32 \times 32$ pixels with a 50\% overlap. The resulting velocity fields have grid spacing of $9.388 \times 10^{-4} \ \text{m} \ (<0.1D)$\rev{,  with dimensions of 89
$\times$ 133 points}. A portion of the dataset used in this work is available and described in further detail in \citet{palmer2025data}.


It is well established that three-dimensional flow physics are present in the range of Reynolds numbers tested \citep{williamson1996review}. Additionally, the primary vortices were observed to be shed at an oblique angle relative to the cylinder; oblique shedding is often attributable to pressure differentials from asymmetries in cylinder end conditions \citep{hammache1991experimental}. While qualitatively similar to parallel vortex shedding, oblique shedding has implications with respect to the shear layer instability and the Strouhal-Reynolds number relationship when compared to parallel shedding \citep{hammache1991experimental, prasad1997three}. Further, in canonical parallel vortex shedding, the shear layer begins to transition at $Re \approx1000$ with the development of secondary streamwise vortices in the recirculation region; in oblique shedding conditions, as studied here, the shear layer instability may be delayed to $Re \approx 2600$ \citep{williamson1996review, prasad1997instability}.  In addition to the challenges associated with increased measurement error from out-of-plane velocities, the non-dimensional time ($t^*$) between Reynolds numbers does not guarantee equivalent Strouhal periods.

\subsection{Triple decomposition}

In the analysis presented, a decomposition of both the measured and predicted velocity fields is used to gain physical insight into surrogate model performance. \rev{This decomposition enables more in-depth analysis and comparison of the learned and observed fluid mechanics across scales.} Given the periodic nature of cylinder wakes, the mean, periodic, and fluctuating components were separated in the velocity fields following \citet{reynolds1971}:
\begin{equation}
u_i = \bar{u}_i + \tilde{u}_i + u'_i.
\end{equation}
Here, $\bar{u}_i$ is the time-averaged mean velocity field, $\tilde{u}_i$ the periodic component, and $u'_i$ represents random fluctuations in the flow. The time-averaged component was calculated by taking the average over all trajectories in the test dataset. 

Proper orthogonal decomposition (POD) was used to estimate the periodic component of the velocity field. POD enables modal decomposition of dynamics onto an orthogonal, \rev{energy-ranked} basis, background and details for which may be found in \citet{berkooz1993pod}. In periodic flows, such as a cylinder wake, the most energetic modes represent the regular periodic fluctuations (e.g., vortex shedding). Reconstructing the velocity fields with only these most energetic modes provides an approximate phase-averaged velocity field representing the periodic component of the flow. 

\begin{figure}
  \centerline{\includegraphics[width=\textwidth]{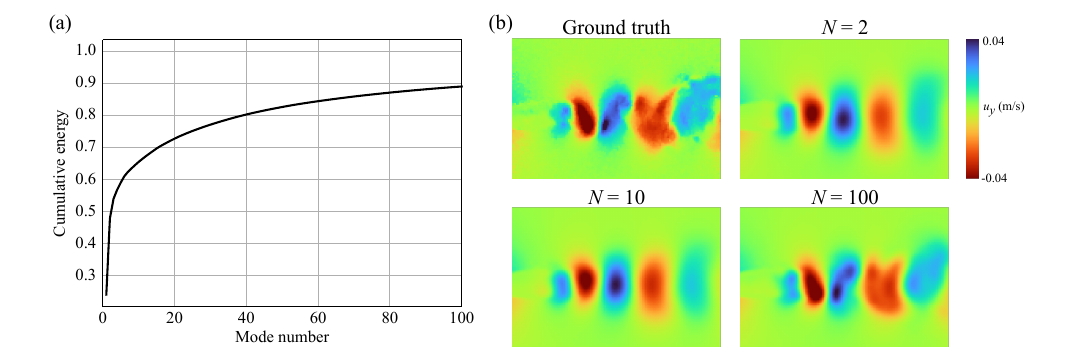}}
  \caption{(a) Cumulative energy of POD modes for $Re=590$. (b) Corresponding velocity field reconstruction with varying number of modes.}
\label{fig:pod}
\end{figure}

The cumulative energy of the first 100 modes and an example velocity field reconstruction can be seen in figure \ref{fig:pod}. This decomposition followed a basic snapshot POD approach. As expected, nearly half the energy is contained in the first two modes alone. The $N=2$ and $N=10$ reconstructions seen in figure \ref{fig:pod}b show qualitatively similar fields, with the $N=100$ case showing more fine-scale spatial detail. In the context of bluff body wakes, the first two modes have been shown to effectively capture dominant vortex-shedding behavior for the purposes of triple decomposition, even at high Reynolds numbers \citep{perrin2007obtaining, oudheusden2005phase}. Following this, as well as the distribution of energy observed, the periodic component of the flow field was estimated using the POD reconstruction from the first two modes alone. This was done separately for all Reynolds numbers and models \rev{to ensure robust decomposition and isolation of the fluctuating component}. 

The irregular fluctuating term ($u'_i$) was then isolated by subtracting both the time-averaged and \rev{estimated} phase-averaged components from the full velocity field. The models were trained to predict the full velocity field rather than any one component from the triple decomposition. The decomposition was exclusively used during analysis. 

\subsection{Surrogate modeling}

In this study, data-driven models are used as surrogates to forecast the time evolution of the $x$- and $y$- component velocity fields in the wake of a circular cylinder at a range \rev{of} Reynolds numbers. Separate models are trained at each Reynolds number to isolate the forecasting performance given the flow conditions at that point. 

Four common data-driven model architectures were chosen for this study to represent a \rev{range of surrogate model modalities}: fully convolutional neural networks (FCNs), U-Nets, Fourier neural operators (FNOs), and dynamic mode decomposition (DMD). U-Nets and FCNs were included to represent common, generalized deep learning frameworks. While U-Nets are often found to surpass FCN performance, FCNs remain commonly deployed for surrogate modeling of fluid flows \citep{achermann2024windseer}. As neural operators, FNOs were included as one of the most common architectures in the growing field of operator learning. DMD, which is not a deep learning method, has been a common modeling tool in the fluid mechanics community for decades and provides important context for the performance of the nonlinear network-based models \citep{schmid2022dynamic}.

At each Reynolds number, three models of each type (i.e., FCN, U-Net, FNO, DMD) were trained with different initializations \rev{and training splits} to gauge mean performance \rev{independent of seeding}. \rev{As no significant or meaningful variation was observed between these trials, only the mean performance was considered in this analysis.}

A simple mean squared error (MSE) loss was used to train the deep learning models, \rev{which were implemented in PyTorch \citep{paszke2019pytorch}}. The deep learning models were trained on a set of 3488 time sequences with a \rev{test set of 832 sequences at each Reynolds number, split randomly and redrawn independently for each of the three runs. The early-stopping criterion was evaluated on the test set.} Each time sequence contains 11 time-steps which are unique to that sequence\rev{, where a time-step represents a single velocity field snapshot} (i.e., sequences had no overlap). The DMD model was trained with a randomly selected subset of this dataset, as discussed below. 

\rev{Prior to training, the velocity field components were standardized separately at each Reynolds number and for each train/test split by subtracting the component mean and dividing by the standard deviation found in the training portion. Separately standardizing the two velocity channels ensured that the loss would not be dominated by the larger-magnitude streamwise velocity component. Model outputs were transformed back to dimensional units using these mean and standard deviation values. No smoothing, masking, or outlier replacement was used when processing the velocity fields. Models operated directly on the two-dimensional, two-component 89 $\times$ 133 grid.}

Using a single measured instantaneous two-dimensional two-component velocity field as input, the data-driven models can recursively predict forward an arbitrary number of time-steps. \rev{In this modality, the first time-step is forecast using the measured observation, with following time-steps being informed by the preceding time-step forecast. This is to say that the first measured field ($\bf{u}(t)$) is passed to the model to predict the subsequent field ($\bf{u^*}(t+1)$). This predicted field $\bf{u^*}(t+1)$ can then be passed as input to the model to predict the following step ($\bf{u^*}(t+2)$), and so on, without further observations.}

During training, the \rev{deep learning} models \rev{(i.e., FCN, U-Net, FNO)} learned to recursively predict ten time-steps. The MSE loss was then calculated over the full \rev{ten-step} forecast horizon \rev{during backpropagation. Physically, this corresponds} to roughly one-third of a full vortex shedding cycle. This multi-step recursive structure improves prediction accuracy over long forecast horizons when compared to training on a single prediction step. \rev{Although only trained to forecast over ten time-steps, the model performance at $Re=590$ was also considered over a held-out and extended thirty time-step horizon to analyze long-term forecast stability.}

Given the limitations associated with experimental quantitative flow visualizations, these instantaneous velocity fields do not fully define the state of the turbulent dynamics. The finite resolution, planar observation, and absence of out-of-plane velocity data all prevent the measurements from observing the full state of the dynamics. This limitation is often mitigated by the use of time-delay embeddings, whereby including a short history of observations as input may better define the system dynamics \citep{karthik2020chaos}. However, this work maintains a single observation input to reflect common limitations of laser repetition rates in many double-pulsed PIV systems, where multiple time-resolved measurements would not be available. A comparison of the deep learning models using a time-delay embedding with a two time-step input is included in the appendix (figure \ref{fig:multistep_error}). While the time-delay embedding slightly improves performance across models, the relative performance remains largely unchanged. It is possible that longer time-delay embeddings may further improve performance, however further investigations in this matter lie beyond the scope of the current work. 

\subsubsection{Fully convolutional networks}

Fully convolutional networks (FCN), introduced by \citet{long2015fully}, are networks comprised entirely of convolutional layers. These layers, commonly used for learning tasks involving images, perform convolution operations between filters and the channels of an input array (i.e., an image). The output of these convolutions is known as an activation map. Through the learning process, the filters change shape to improve activation relative to local features contained in the data. Each filter in a convolutional layer corresponds to a channel in the output of that layer, typically resulting in the output shape having less spatial resolution and more channels. 

Here, we applied an FCN with an ``encoder-decoder" structure. In this case, the subsequent outputs of the first several layers form an encoder, and have progressively less spatial resolution and more channels. We perform additional convolutions on the latent representation following the encoder portion, where the spatial information of the initial image is most dense. This is followed by a decoder portion, where the output can be returned to the desired dimension through a series of transposed convolutional layers. The hyperparameters used for FCN can be found in the appendix (table \ref{tab:fcnparam}).

\subsubsection{U-Net}

One of the most popular network architectures in modern deep learning, U-Net was first introduced by \citet{ronneberger2015unet} for biomedical image segmentation. U-Nets can be structured with convolutional layers similar to FCNs, but also feature skip connections where the output of several layers of the encoder portion of the network \rev{is} directly concatenated onto the input of corresponding layers in the decoder portion of the network. This helps the network maintain spatial information that may have been otherwise lost at the bottleneck between encoder and decoder. The hyperparameters used for U-Net can be found in the appendix (table \ref{tab:unetparam}).

\subsubsection{Fourier neural operators}

Neural operators are distinct from standard neural networks (e.g., FCNs and U-Nets) for their ability to directly approximate operators from data alone \citep{kovachki2021neural}. This includes operators that map between sets of functions related by a family of PDEs (e.g., the Navier-Stokes equations), making these methods of particular interest for problems in fluid mechanics.

Introduced by \citet{li2021fno}, Fourier neural operators (FNOs) use unique Fourier layers to achieve guaranteed universal approximation for continuous operators \citep{kovachki2021fno}. These Fourier layers consist of two paths: non-linear activation functions as found in classical neural networks, and a linear transform of the input signal performed in Fourier space. The non-linear activation function path serves to approximate local nonlinearities, and the Fourier transform path serves as a global integral operator to account for non-local effects in real space. The paths are combined when passed forward, and the networks are trained using the same principles as neural networks. The hyperparameters used for FNO can be found in the appendix (table \ref{tab:fnoparam}).

\subsubsection{Dynamic mode decomposition}

Dynamic mode decomposition (DMD), introduced by \citet{schmid2010dynamic}, is a method of data-driven reduced-order modeling. A departure from the network-based deep learning methods discussed above, DMD allows for the development of reduced-order linear models of nonlinear dynamical systems.  Especially popular in fluid mechanics applications, DMD enables the numerical estimation of a Koopman operator approximation, which is typically formulated to represent the evolution of a dynamical system by a single time-step. For this work, we used a variant of DMD introduced by \citet{heas2022low} \rev{as implemented in the PyDMD library} with SVD factorization of rank 250 \citep{ichinaga2024pydmd}. The specific rank was chosen through \rev{a parameter sweep}; higher rank factorizations saw diminishing returns. \rev{Similarly, a sweep across training-set sizes found DMD model performance to be insensitive to sample count beyond approximately 500 sequential time-step pairs. Following this, all DMD models used in this work were trained with an equivalent of 2000 sequential time-step pairs (i.e., equivalent to 200 time sequences). Despite the relatively large rank and dimensionality of the dataset, the dynamic mode decomposition model is comprised of fewer than 6 million parameters, which is approximately 10\% of the number of parameters defining the deep learning models presented in this work.}

\subsubsection{Evaluation speed}
\begin{table}[b]
\caption{\label{tab:eval_speeds}Average model evaluation time on an NVIDIA A100 \rev{for batch size of one}. All models evaluate faster than the minimum inter-sample time (5 ms). Note that \rev{inference time is independent of Reynolds number and the} deep learning models may be evaluated even faster if compiled.}
\begin{ruledtabular}
  \begin{center}
\def~{\hphantom{0}}
  \begin{tabular}{lc}
  \textrm{Model} & \textrm{Evaluation Time} \\
  \colrule
U-Net & 2.52  ms\\
FNO & 3.13  ms\\
DMD & 1.31  ms\\
FCN  & 2.02 ms \\
  \end{tabular}
  \end{center}
\end{ruledtabular}
\end{table}

Table \ref{tab:eval_speeds} shows the respective mean evaluation times of the models. The velocity fields are predicted in increments of non-dimensional time approximately equal to \rev{$t^* \approx 0.16$}. As the non-dimensional time was held constant, the corresponding physical time between snapshots varies with Reynolds number, from $5$ ms ($Re = 2920) $ to $62.5$ ms ($Re=230$). Given a minimum inter-sample time of $5$ ms ($Re = 2920$), table \ref{tab:eval_speeds} shows that all surrogate models tested are capable of forecasting \rev{velocity fields} faster-than-real-time \rev{for the data presented here. While these times only reflect model inference, modern real-time rapid PIV systems are capable of processing over 1000 frames-per-second with low-latency}, potentially enabling future model-based predictive flow control applications \citep{bollt2025piv}.

\section{Results}\label{sec:results}

This section presents an analysis of data-driven surrogate models, namely U-Net, FNO, DMD, and FCN, in the context of time-forecasting experimentally measured turbulent cylinder wakes. We first examine performance  across extended forecast horizons at a fixed Reynolds number ($Re=590$), comparing the temporal dynamics of the fluid flow in predicted and experimentally-measured velocity fields. Following this we explore how surrogate models contend with increasingly turbulent dynamics by characterizing performance across Reynolds numbers ranging from the start of the three-dimensional wake transition ($Re=230$) to \rev{shear} layer transition ($Re=2920$). In addition to quantifying performance relative to flow physics, we examine the impact of three-dimensional turbulent fluctuations in the context of observing and measuring the state of the system. Finally, the model performance is considered simultaneously across Reynolds numbers and forecast horizons. 

\subsection{Propagating fine-scale fluid dynamics over extended forecast horizons}

\begin{figure}
  \centerline{\includegraphics[width=\textwidth]{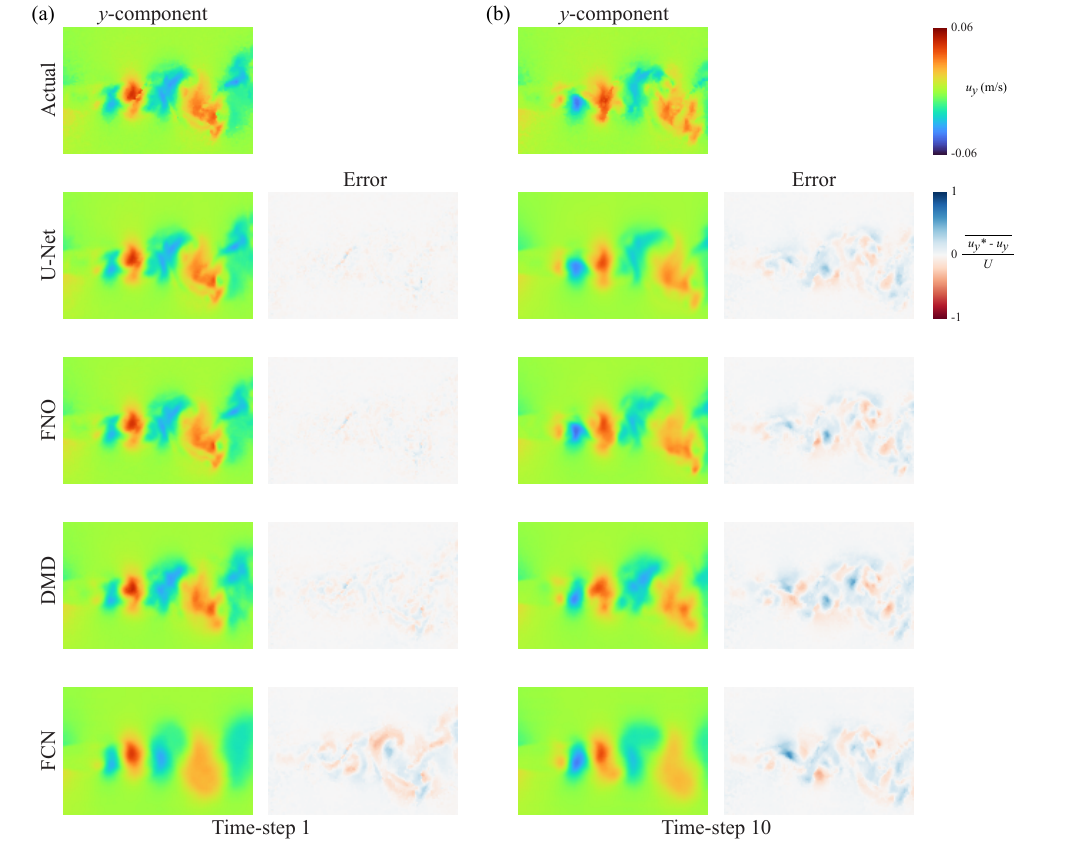}}
  \caption{Actual and predicted y-component velocity fields with error for all models. The results of the (a) first and (b) tenth recursive prediction for $Re = 590$ are shown.}
\label{fig:error_snapshots}
\end{figure}

While recursive application can enable arbitrary forecast horizons for surrogate models, compounding errors can limit the utility of and distort dynamics in long prediction sequences. Figure \ref{fig:error_snapshots} shows example predictions and corresponding error fields of the \rev{transverse} velocity component for all four model types at the first and tenth step in a recursive forecast at a fixed Reynolds number of $Re=590$.

Generated directly from the experimental measurements, the predictions at the first time-step (figure \ref{fig:error_snapshots}a) generally show low error and a high level of spatial detail. However, by the tenth time-step (figure \ref{fig:error_snapshots}b), the forecast velocity fields appear significantly less detailed than the actual field, and more significant error has accumulated. The error in the velocity fields appears high frequency, which is consistent with the apparent smoothness of the predictions. 

The FCN surrogate is a notable exception to this pattern, with the first time-step prediction already appearing quite smooth with high error, and the tenth time-step prediction appearing qualitatively similar to the first. 

\begin{figure}
  \centerline{\includegraphics[width=\textwidth]{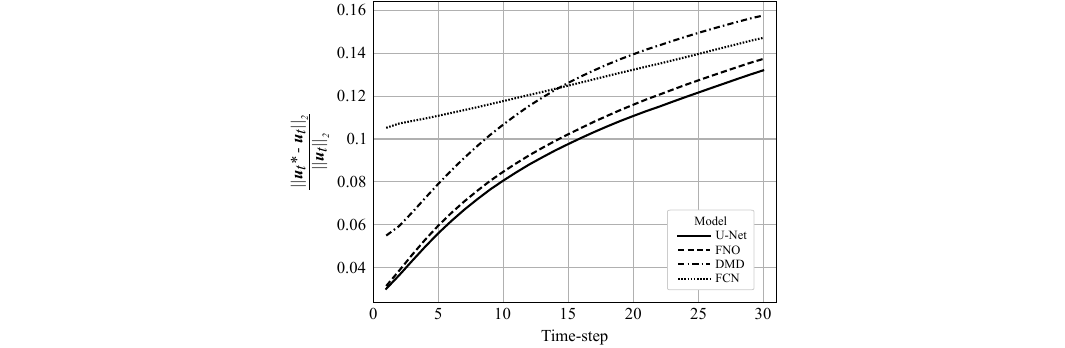}}
  \caption{Mean error per time-step for various models at $Re = 590$. \rev{A comparison with the persistence baseline can be found in figure \ref{fig:error_vs_time_pers}.}}
\label{fig:error_vs_time}
\end{figure}

Quantifying the performance as a function of forecast length, figure \ref{fig:error_vs_time} shows the normalized error of the four models at the same Reynolds number ($Re=590$) across a thirty time-step horizon. This horizon was chosen as it is approximately equivalent to a full vortex shedding cycle. \rev{For additional context, an equivalent plot including the persistence prediction error (i.e., difference between actual field and original input) can be found in figure \ref{fig:error_vs_time_pers}. Notably, the FCN prediction error is actually greater than the persistence baseline over the first several time-steps. This is consistent with the immediate over-smoothing observed for the FCN model in figure \ref{fig:error_snapshots}.} 

As shown in figure \ref{fig:error_vs_time}, the error accumulates most rapidly over the first several prediction time-steps for the U-Net, FNO, and DMD models. However repeated recursive application of the surrogate model does not accelerate error accumulation but rather stabilizes it, as indicated by the reduction in error rate-of-change at longer time horizons. The FCN forecast is an outlier from this trend, with a large initial error and a moderate linear increase in error across the forecast horizon. As a result, the FCN surrogates show the lowest change in error per time-step for the majority of the prediction horizon, and achieve a lower absolute error than the DMD surrogate models after 14 prediction time-steps.

While the FCN surrogate models appear as an outlier, their performance provides important insights into the abilities of surrogate models in general, especially with respect to experimental measurements. The high initial value and linear increase in error are consistent with the overly smooth predictions produced by the FCN model in figure \ref{fig:error_snapshots}, suggesting that FCN surrogates learn to avoid spatial detail and predict only the dominant vortex shedding pattern. This is unsurprising; the FCN is effectively structured as an autoencoder, with the velocity field reduced to a lower dimensional representation and passed through the bottleneck before reconstruction. The FCN is therefore poorly suited for preserving spatial detail, with MSE loss pushing the weights to preserve only the most dominant, low-frequency dynamics in the reduced state at the bottleneck. 

While achieving lower error forecast predictions, similar challenges apply to the U-Net and FNO surrogate models.  It is unsurprising that U-Net and FNO are better at modeling high-frequency dynamics; the skip-connections that differentiate the U-Net and FCN help preserve spatial details, and the Fourier layers in FNOs transform a portion of the signal to Fourier space wherein linear operations serve as global integrators to capture non-local effects \citep{ronneberger2015unet, li2021fno}. However, the relative smoothness of the predictions in figure \ref{fig:error_snapshots}b \rev{shows that high-frequency spatial details are still lost after several prediction steps. This may suggest that the FCN, U-Net, and FNO surrogates suffer from a spectral bias. This is a known phenomenon wherein neural networks prioritize low-frequency features, acting   as a barrier to learning high-frequency components \citep{rahaman2019spectral}. However, this smoothness may also be inherent to the learning problem itself. The optimal predictor under mean squared error given partial observations is the conditional mean \citep{bishop2006pattern, chorin1998optimal}. Given the limited resolution and dimensionality of the observation, some smoothing may be the result of incomplete information regarding the state of the dynamics at high spatial frequencies. It should also be noted that, as linear models, the DMD-based surrogates cannot exactly represent the conditional mean of the nonlinear dynamics, regardless of the completeness of state observations.}

This \rev{smoothing} is also consistent with the decreasing rate-of-change in error accumulation in figure \ref{fig:error_vs_time}. As the high-frequency component decays with recursive predictions, the increase in error per time-step becomes dominated by a change in error for the predicted mean or periodic components of the velocity field towards which the models may be biased. To better understand and analyze the limitations of physical dynamics learned by the various surrogates through incomplete observations (i.e., 2D-2C velocity fields) \rev{and understand the mechanical basis for this smoothing}, we must isolate the specific features driving error. 

\begin{figure}
\centerline{\includegraphics[width=\textwidth]{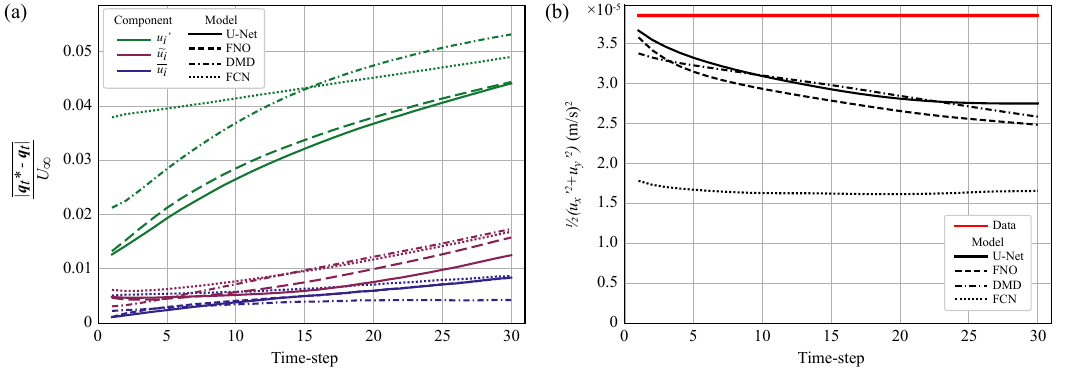}}
  \caption{\rev{Decomposed velocity and estimated planar TKE for $Re=590$ case averaged over the flow field.} (a) Comparison of fluctuating, periodic, and mean component error. (b) Estimated mean planar turbulent kinetic energy in model forecast and measurements across time-steps.}
\label{fig:fluc_errors_tkes_time}
\end{figure}

As detailed in the methods section, a triple-decomposition was performed to isolate the fluctuating component of the velocity field, absent of the periodic and time-averaged component. The periodic component is approximated by projecting the velocity field onto the two most energetic modes found through POD \citep{perrin2007obtaining, oudheusden2005phase}.

Figure \ref{fig:fluc_errors_tkes_time}a shows the error of the fluctuating, periodic, and time-averaged components of the velocity field as a function of prediction step. Note that the errors are normalized by the magnitude of the free-stream velocity (rather than the norm of the field as in figure \ref{fig:error_vs_time}). The error is dominated by that of the irregular velocity fluctuations. Further, the shape of the fluctuating component error with respect to time-step matches figure \ref{fig:error_vs_time} very closely. This indicates that these unsteady fluctuations are the primary driver of error accumulation.

The fluctuating velocity component can be used to estimate the \rev{mean} planar turbulent kinetic energy (TKE) in the predictions, as shown in figure \ref{fig:fluc_errors_tkes_time}b as a function of the prediction time-step. Given that the experimental conditions were held constant and the time trajectories are not phase-aligned, the experimentally measured estimate for mean planar TKE is approximately constant across the forecast horizon, as represented by the red line. The standard deviation of the \rev{measured value} is also plotted, but is occluded by the thickness of the line itself. 

None of the surrogate models preserve the estimated planar TKE across the forecast horizon. The FCN surrogate model produces forecasts with greatly reduced TKE at the first time-step followed by only a small decay and ensuing plateau; this should be expected given the lack of high-frequency dynamics seen in figure \ref{fig:error_snapshots}. The estimated planar TKE in U-Net and FNO surrogate predictions decays nonlinearly, with the most rapid decrease happening over the first few time-steps. The \rev{decay of TKE} in the U-Net surrogate model prediction appears to slow at around the 25th time-step, indicating that it may indefinitely maintain considerably more TKE than the FCN surrogates. This is likely attributable to the skip connections that differentiate the two models, indicating that \rev{these detail-preserving connections} are  critical in learning and maintaining turbulent energy in the fluctuating component of the velocity field. 


While the DMD surrogate model results were qualitatively similar to those for U-Net and FNO in figures \ref{fig:error_snapshots}, \ref{fig:error_vs_time}, and \ref{fig:fluc_errors_tkes_time}a, differences in the estimated planar TKE (figure \ref{fig:fluc_errors_tkes_time}b) reflect fundamental differences between these model types. The prior figures indicate the same trends of progressive smoothing at higher time horizons for DMD, which were explained for the other models by spectral bias \rev{and limited observability}. However, the DMD-based surrogate models are not neural networks, and therefore are not subject to spectral bias. Rather, in DMD the most energetic modes are explicitly prioritized through the singular value decomposition used to compose these linear models. Instead, we turn to the estimated planar TKE for an explanation. 

The decay in estimated planar TKE for the DMD surrogate model forecasts appears linear with more energy content than other model predictions across most of the prediction horizon. This linear decrease can be attributed to another interaction between experimental data and data-driven surrogate models; observation noise can result in underestimated eigenvalue magnitudes for the model, causing a deviation in growth or decay rates \citep{duke2012error, hemati2017biasing}. Given the presence of experimental noise and unobserved three-dimensional dynamics, the linear model predicts a small decrease in energy for each time-step which results in the linear decay seen across the forecast horizon in figure \ref{fig:fluc_errors_tkes_time}b. As mentioned, this results in the DMD forecast velocity fields having the highest TKE for the majority of the time-steps shown; however, considering the error of the fluctuating component seen in figure \ref{fig:fluc_errors_tkes_time}a, it is worth noting that this preservation of TKE does not result in an overall reduction in prediction error. 

\rev{Some portion of the TKE deficit seen in figure \ref{fig:fluc_errors_tkes_time}b may reflect measurement noise, which all the presented models are unable to reproduce. However, as the measurement noise is stationary across the prediction horizon, its contribution is bounded by the deficit at the first prediction step and it cannot account for the progressive decay of estimated planar TKE. The gradual smoothing and decay of TKE is further supported by directly examining the spatial energy spectra at various prediction time-steps, as found in figure \ref{fig:energy_spectra}. The energy content matches well at low wavenumbers across the prediction horizon, but a clear deficit emerges at higher frequencies for all models. } 

While the model performances vary, several trends emerge which are relevant for more general surrogate model forecasting of experimental or real-world fluid flows. First, error tends to increase most rapidly over the first prediction time-steps. Figure \ref{fig:error_vs_time} shows that the normalized error more than doubles for the best-performing model (U-Net) across the first ten time-steps. Qualitatively, this corresponds to spatially smoother and less energetic predictions which are attributable in-part to the inherent spectral bias of neural networks \rev{(for the network-based surrogates)}. This \rev{is} exacerbated further by experimental noise and unobservable out-of-plane or sub-grid dynamics, \rev{under which the optimal predictor converges to the conditional mean}. Given an inadequately defined state observation, models cannot effectively learn to propagate the dynamics indefinitely, and instead produce smooth predictions with high-frequency features observed from the initial input decaying in time. While the practical utility of long-term predictions is highly context dependent, none of the models tested appear capable of accurately predicting or preserving the development and propagation of high-frequency velocity components for more than a few time-steps. Over shorter prediction horizons (e.g., $<5 $ time-steps, approximately 1/6th of a vortex shedding cycle), however, we see that models appear to maintain some existing velocity fluctuation features in predictions despite the challenges associated with experimental measurements and limited observability. 



\subsection{Learned high-frequency dynamics across Reynolds numbers}

Looking at a single Reynolds number, the relationship between flow physics and surrogate model performance has become more clear. Next we can examine how these dynamics change across Reynolds number. There exist multiple challenges associated with varying the Reynolds number in the context of modeling partially observed experimental velocity fields, which cannot be uncoupled. As the Reynolds number increases, turbulent fluctuations become more energetic and three-dimensional. Irregular out-of-plane velocity fluctuations at large Reynolds numbers are especially problematic; they cannot be measured with the planar two-component PIV, and they actively diminish the fidelity of in-plane measurements by moving seed particles in and out of the laser sheet. Out-of-plane motion of seed particles through the laser sheet results in reduced correlation between image pairs and more noise in the resulting velocity fields. As a result, increasing Reynolds number means more irregular fluctuations, more energy outside of the observable state, and noisier measurements.

To better understand how these characteristics impact the performance of data-driven surrogate models, separate models were trained on experimental datasets at seven Reynolds numbers ranging from $Re=230$ to $Re=2920$.

\begin{figure}
\centerline{\includegraphics[width=\textwidth]{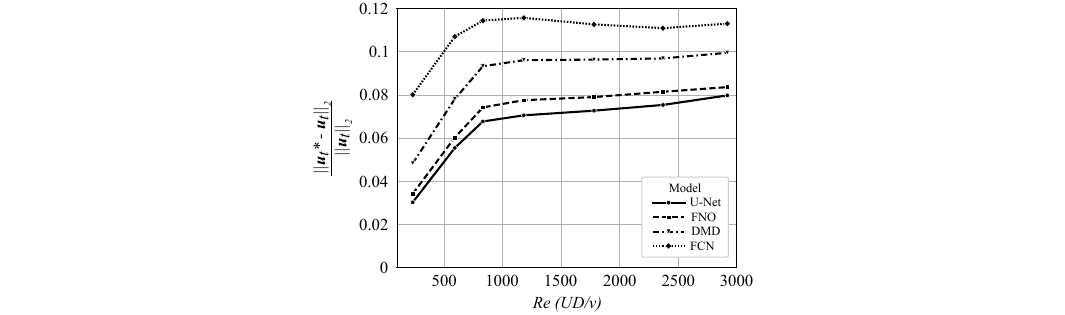}}
  \caption{Relationship between mean prediction error and Reynolds number for all models. }
\label{fig:reynolds}
\end{figure}

Figure \ref{fig:reynolds} shows the mean prediction error as a function of the Reynolds number averaged across a ten time-step prediction horizon. It appears that the performance is most sensitive to Reynolds number from $Re=230$ to $Re=830$ for all four models, where there exists a rapid increase of error with respect to Reynolds number. The performance for all four model types changes more gradually at Reynolds numbers greater than $Re=830$, with the error for FCN surrogates actually decreasing slightly between $Re=1180$ and $Re=2370$. The challenges associated with data-driven modeling at increasing Reynolds number (e.g., turbulent fluctuations, out-of-plane velocities, energetic sub-grid eddies, experimental noise) do not appear to result in continually increasing error at $Re \ge 830$.

\begin{figure}
  \centerline{\includegraphics[width=\textwidth]{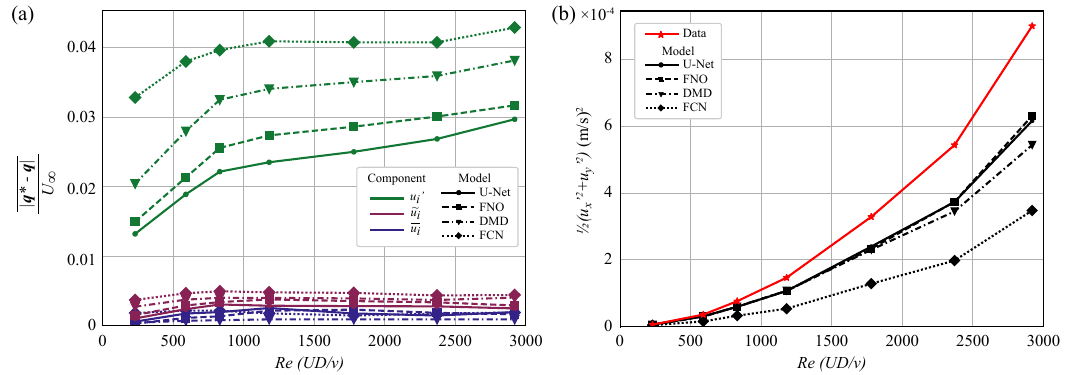}}
  \caption{(a) Comparison of fluctuating, periodic, and mean component error. (b) Estimated planar turbulent kinetic energy \rev{averaged over the flow field} in model forecast and measurements  across Reynolds number.}
\label{fig:tkes_decomposed_err}
\end{figure}

Figure \ref{fig:tkes_decomposed_err}a decomposes the error into the fluctuating, periodic, and time-averaged velocity components, normalized by the free-stream velocity. As in figure \ref{fig:fluc_errors_tkes_time}a, the error is dominated by the fluctuating velocity component. While the trend with respect to Reynolds number seen in figure \ref{fig:tkes_decomposed_err}a is similar to the trend seen in figure \ref{fig:reynolds}, the U-Net, FNO, and DMD fluctuating velocity errors appear more sensitive to Reynolds number than the full field normalized error. 

Figure \ref{fig:tkes_decomposed_err}b shows the estimated \rev{mean} planar TKE contained in the two-dimensional two-component velocity fluctuations \rev{in} the measured field as well as model forecast predictions averaged across a ten time-step horizon. The surrogate models tend to underestimate the planar TKE, with the gap growing with increasing Reynolds number. This gap is largest across Reynolds numbers for FCN; based on observations regarding FCN performance in the previous section, it is unsurprising that it remains an outlier with much lower estimated planar TKE than the other models across the Reynolds numbers shown. 

Notably, the predicted TKE for DMD forecasts is close to that of FNO and U-Net despite having markedly higher error in the fluctuating velocity component. As discussed in the previous section, DMD-based surrogate models explicitly include the highest energy modes. The decrease in TKE for the DMD surrogate model seen at $Re=2920$ may be the result of increased experimental noise reducing the magnitude of eigenvalues further.

\begin{figure}
  \centerline{\includegraphics[width=\textwidth]{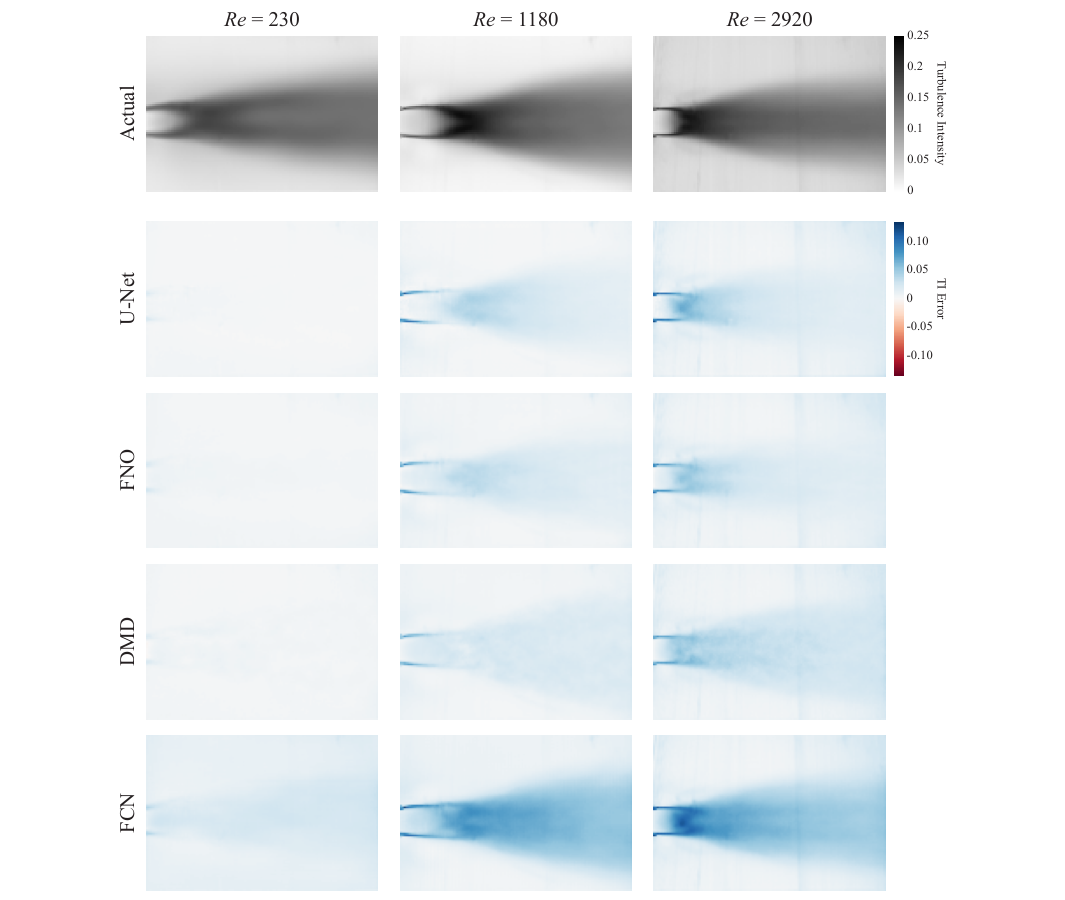}}
  \caption{Measured turbulence intensity ($TI = \frac{\sqrt{u_x'^2 + u_y'^2}}{2 U_{\infty}}$) in the cylinder wake at $Re=230$, $Re=1180$, and $Re=2920$, as well as the corresponding mean error fields for turbulence intensity in surrogate model forecasts. The error is calculated as the difference between the actual and predicted turbulence intensity fields. Positive error values suggest models tend to underestimate turbulence intensity.}
\label{fig:turbulence_intensity_error_compare}
\end{figure}

Figure \ref{fig:turbulence_intensity_error_compare} provides physical insight into the error accumulated by showing the estimated turbulence intensity derived from the two-dimensional TKE, along with the error in turbulence intensity for models at three Reynolds numbers. Here the turbulence intensity was calculated from the turbulent kinetic energy averaged across the ten time-step forecast horizon for all candidates in the testing dataset. As expected from figure \ref{fig:tkes_decomposed_err}b, the surrogate models tend to underestimate the turbulence intensity throughout the region of interest as indicated by the positive turbulence intensity error. 

At the lowest Reynolds number shown ($Re=230$), there is little error for U-Net, DMD, and FNO surrogate models, except at the leading portion of the shear layer surrounding the recirculation region. The FCN forecasts underestimate the turbulence intensity at this Reynolds number by a wider margin, even in the free-stream. This is consistent with the smooth predictions produced by FCN previously noted at $Re=590$ (figure \ref{fig:error_snapshots}), which appears to continue even for the lower Reynolds number case shown here.   

At the intermediate Reynolds number shown, $Re=1180$, the models underestimate the turbulence intensity in the full length of shear layer surrounding the recirculation region. Again, the FCN predictions show the largest deviations in turbulence intensity, with the distribution of error closely following the shape of the measured turbulence intensity field. Given the characteristics of FCN surrogate model predictions previously observed, it is unsurprising that the turbulence intensity is underestimated throughout the field. The other two deep learning based surrogate models (U-Net and FNO) are similar with respect to the turbulence intensity error fields at $Re=1180$, with error primarily accumulating near the vortex formation region. This can likely be attributed to velocity fluctuations during the vortex formation process and the resulting irregularly formed vortices, or even vortex dislocation events. While the turbulence intensity error field for DMD surrogate models appears to have similar magnitude to FNO and U-Net, it is more uniformly distributed throughout the region where vortices are formed and convect downstream. This distribution may be the result of the nature of DMD, where we would expect errors to accumulate from high-frequency, low energy transient features rather than primary vortex formation. 

At the highest Reynolds number tested, $Re=2920$, experimental artifacts in the training data become apparent. Nearly vertical imperfections appear in the measured turbulence intensity, which can be attributed to uneven power distribution in the laser sheet arising from defects in the optical surface at the bottom of the test section. These artifacts become apparent at the highest Reynolds numbers in the dataset due to increased frame rate and subsequent reduced exposure time. The models do not predict the artifacts, which is why they appear as vertical lines in figure \ref{fig:turbulence_intensity_error_compare}. Looking past the artifacts, the comparative performance of the models is similar to the $Re=1180$ case. Notably, this is the only Reynolds number tested which lies firmly within the shear layer transition regime for oblique shedding \citep{prasad1997instability}. The resulting instability can result in secondary vortices forming in the recirculation region, upstream of the primary vortex formation region. This may be the cause of the accumulated error observed between the shear layers upstream of the vortex formation region for all four models, where secondary streamwise vortices begin to form during the transition. 

\begin{figure}
  \centerline{\includegraphics[width=\textwidth]{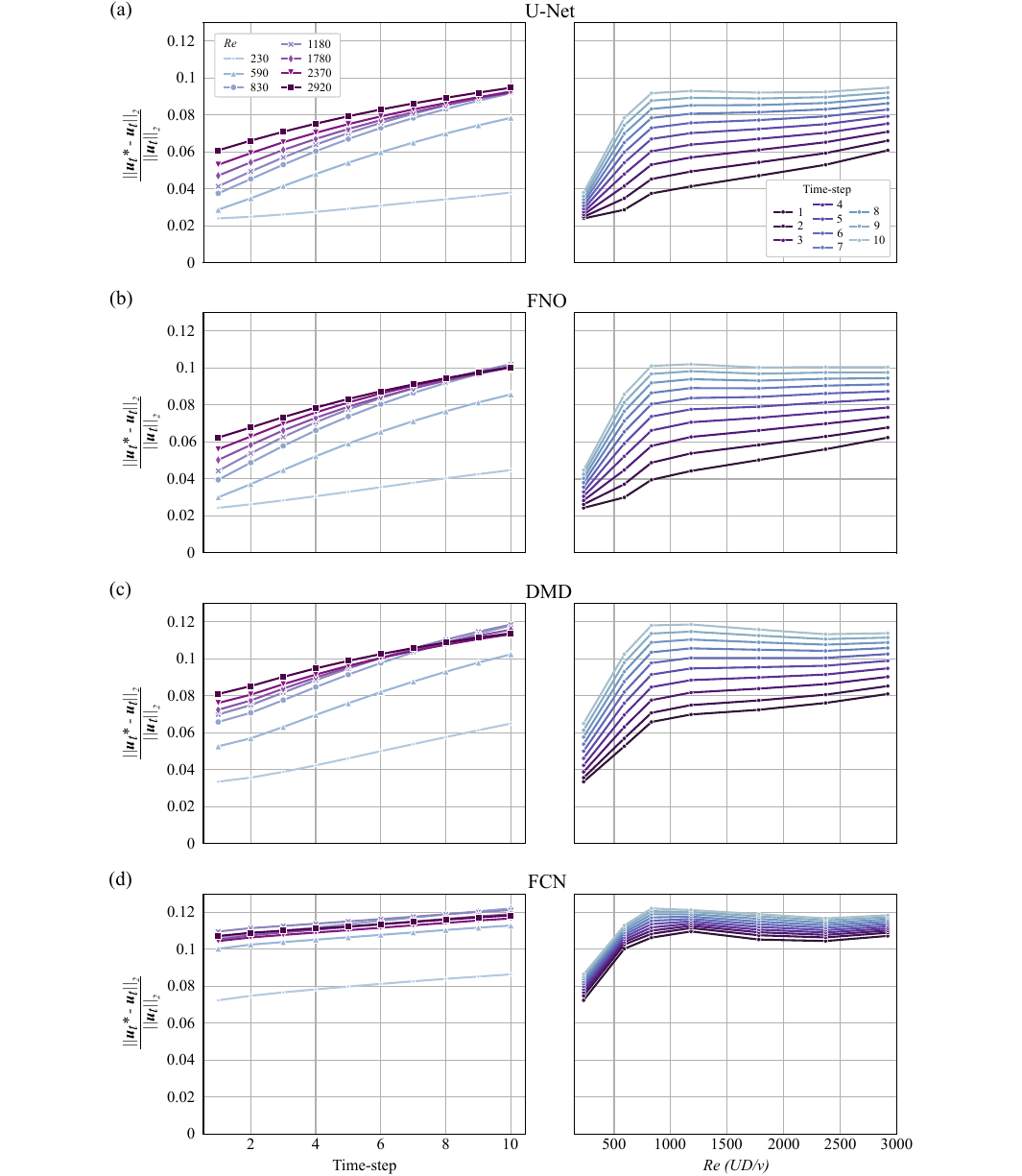}}
  \caption{Relationships between error, time-step, and Reynolds number for (a) U-Net, (b) FNO, (c) DMD, and (d) FCN models. On the left, the error is plotted as a function of prediction time-step across Reynolds number tested. On the right, the error at each time-step is plotted.}
\label{fig:time_and_reynolds}
\end{figure}

The deaggregated error results across both forecast horizon and Reynolds number are shown in figure \ref{fig:time_and_reynolds}. On the left of figure \ref{fig:time_and_reynolds}, normalized error for each tested Reynolds number is plotted against time-step. On the right of figure \ref{fig:time_and_reynolds}, the normalized error for each time-step is plotted against Reynolds number. Broadly, these results recapitulate the general trends observed in figures \ref{fig:error_vs_time} and \ref{fig:reynolds}. However, decomposing time-step and Reynolds number separately helps elucidate specific differences in model performance across the testing regime and provides insights into the physical mechanisms underlying error accumulation. 

Focusing first on the left hand side of figure \ref{fig:time_and_reynolds} provides important context regarding surrogate model performance across Reynolds numbers. For U-Net, FNO, and DMD-based surrogates, the lowest Reynolds number case ($Re=230$) is a clear outlier with comparatively lower absolute error and error rate of accumulation. Notably, looking at these same models, the maximum rate of change in error accumulation per time-step occurs at $Re=830$. Although the error at the first time-step continues to increase for these models at higher Reynolds numbers, the error accumulated at each additional time-step actually decreases. Consequently, the error at the tenth time-step for both U-Net and FNO approximately converge for all Reynolds numbers greater than $Re=590$. A similar trend holds for DMD, with the error for the higher Reynolds numbers intersecting around the seventh time-step. At the tenth prediction step for DMD, the mean normalized error at $Re=830$ is higher than for $Re=2920$. This may indicate that the noisier data and resulting lower-magnitude eigenvalues actually decrease absolute error by producing less detailed predictions at even moderate forecast time horizons. The FCN models are distinct, with similarly high initial error for all but the lowest Reynolds number case, and only a small amount of additional error accumulated at each time-step. This is consistent with the smoothness and relatively high error signal observed in figure \ref{fig:error_snapshots}.

The right side of figure \ref{fig:time_and_reynolds} shows how the error at a given time-step varies across Reynolds numbers. With the exception of the FCN surrogates, it appears that error increases monotonically with Reynolds number at the first prediction steps. However, by the eighth time-step, the error reaches a clear plateau at $Re=830$, similar to the trend seen in figure \ref{fig:reynolds}. The FCN surrogate model predictions show a similar shape across all time-steps, where the error does not grow with Reynolds number beyond $Re=830$.

The sensitivity to Reynolds number seen for the first time-step for U-Net, FNO, and DMD-based models is consistent with the observations in figure \ref{fig:tkes_decomposed_err}b: as the Reynolds number increases, so too does the difference between measured and predicted TKE. For U-Net and FNO specifically, the increasing challenges associated with \rev{incomplete} state observations (e.g., experimental noise, limited dimensionality of observations) and the spectral bias of neural networks both lead to predictions with a reduction in high-frequency content when compared to the physical velocity fields. While these fluctuations become increasingly energetic with Reynolds number, the simultaneously more restrictive limitations in observability prevent the model from effectively learning the dynamics. Unable to propagate or forecast fine-scale dynamics, the surrogate models for $Re\ge830$ collapse onto smooth predictions that minimize MSE. As a result, the first prediction step acts to truncate the high-frequency content and therefore appears most sensitive to Reynolds number. The subsequent time-steps lead to increasingly smooth predictions and less sensitivity to Reynolds number.

\section{Conclusion}\label{sec:conclusion}

In this work, we have analyzed an application of data-driven surrogate models for predicting experimentally measured fluid flows. These models have the potential for faster-than-real-time evaluation, possibly enabling real-world deployment for predictive model-based flow control. Common surrogate model architectures such as U-Net, FNO, DMD, and FCN were evaluated and their predictions analyzed in the context of observability limitations and increasingly complicated flow physics. The data-driven models were trained and evaluated on velocity fields measured in the near wake of a circular cylinder at a range of Reynolds numbers. The state observations were measured through two-dimensional, two-component PIV, which has limited resolution and explicitly excludes out-of-plane velocity components which are present throughout the range of Reynolds numbers tested. 

We find that data-driven models are able to propagate low-frequency dynamics over long forecast horizons tested at $Re=590$, however fine high-frequency spatial details decay rapidly over the first several prediction time-steps. The preservation of high-frequency flow features was significantly improved by the presence of spatial-detail preserving network features (e.g., skip-connections). Still, through a triple decomposition, we established that the majority of the error comes from the high-frequency fluctuating velocity component. We see this in the decay of estimated planar TKE shown for all four models in figure \ref{fig:fluc_errors_tkes_time}b. \rev{This is also seen directly in the spatial energy spectra (figure \ref{fig:energy_spectra}), where the energy matches well across prediction horizons for low wavenumbers while high-frequency content decays.} While the goal of this work is not to provide a benchmark comparison between the models, we can also note that the highly generalized U-Net architecture achieved the lowest error in all metrics shown (with the exception of TKE magnitude) across the forecast horizon. However, linear DMD surrogate models better preserved estimated planar TKE at intermediate forecast horizons while requiring fewer parameters and \rev{using} significantly less \rev{training} data.

With respect to Reynolds number, we find the error across data-driven surrogates to have two distinct regimes, with mean forecast error being notably more sensitive to Reynolds number at $Re<830$ than for $Re=830$ to $Re=2920$ (figure \ref{fig:reynolds}). This is attributable to a combination of factors, including the increasingly unsteady and turbulent dynamics which present a greater learning challenge, and the increase in measurement noise from the associated rising three-dimensional effects. However, rather than the growing challenge leading to a corresponding growth in error, the effective result is more relative smoothing in the predicted velocity fields. This smoothing effectively truncates high-frequency dynamics making the resulting error less sensitive to Reynolds number. 

While the very presence of fluctuating three-dimensional dynamics implies that the two-dimensional, two-component velocity measurements used in training and testing are insufficient for fully observing the flow field, we found that models can still accurately predict the large-scale dynamics across short time-horizons for all Reynolds numbers tested. Given the chaotic nature of the turbulent dynamics, it seems unlikely that any surrogate would be capable of accurately forecasting the development and evolution of secondary instabilities and finer scale flow features across moderate time horizons, although it remains possible that such a model may match the dynamics in a statistical sense.

We note that this work shows performance only of common data-driven models in baseline configurations; there are surely variants and modifications which could achieve superior prediction accuracy and help resolve some of the specific problems mentioned. For example, recently \citet{karn2026advno} demonstrated how generative models can help overcome spectral bias in neural networks for fluid mechanics. Additionally, physics-informed learning constraints could further enhance model performance \rev{and help overcome limited and incomplete observations} \citep{li2021pi, karniadakis2021physics}. However, \rev{physics-informed} losses must be approached carefully when working with experimental data; assumptions of divergence-free velocity fields, for example, would be invalid for models trained on the data presented in this work given the unknown out-of-plane velocity components.

This work represents a step towards applications of learning for the forecasting of real-world fluid mechanics. Data-driven models, in conjunction with \rev{high-speed} real-time PIV systems, could enable state-aware predictive flow control \citep{willert2010real,kreizer2010real,varon2019adaptive,bollt2025piv}. \rev{While the models applied in this work represent simplified baselines, practical applications would likely feature more complicated architectures generalizable across Reynolds numbers and flow conditions. Conditioning the model on the observed free-stream velocity could help future models address these challenges. Further,} the \rev{limited} two-component two-dimensional velocity field information is impractical in most real-world engineering applications. However, flow-field reconstruction from sparse measurements has been a focus of many recent works \citep{bright2013,callaham2019robust,loiseau2018sparse,fukami2022machine,fukami2021global,erichson2020shallow,manohar2022sparse}. These reconstruction methods could be used directly in conjunction with the data-driven methods applied here to predict the future state of full-field data from a set of distributed sparse sensors. 

\section{Acknowledgments}
This work was supported by the National Science Foundation Graduate Research Fellowship under Grant No. DGE‐1745301 and the Center for Autonomous Systems and Technologies at Caltech. The authors would also like to thank Alejandro Stefan Zavala for helpful discussions.

\section{Author Contributions}
P.R. and M.G. conceptualized this work. P.R. and C.W. collected the data for this work. P.R. developed and ran models. E.P. and P.R. analyzed data and generated visualizations. P.R. drafted manuscript. All authors reviewed and edited. M.G. supervised. 

\section{Data Availability}
The data that supports the findings of this study are available in \citet{palmer2025data}.

\appendix
\setcounter{figure}{0} 

\renewcommand{\thefigure}{\thesection.\arabic{figure}}

\section{Hyperparameters and Additional Figures}\label{appB}

\begin{table}
\caption{\label{tab:fcnparam}FCN Hyperparameters}
\begin{ruledtabular}
\begin{tabular}{lc}
Initial learning rate  & $10^{-3}$ \\
Max epochs & 200 \\
Batch size & 32 \\
Train set size & 3488 \\
Test set size & 832 \\
Optimizer & AdamW \\
Learning rate scheduler & CosineAnnealingLR w/ warmup \\
Learning rate scheduler warmup epochs & 5 \\
Weight decay & $10^{-4}$ \\
Gradient clipping & 0.50 \\
Early stopping patience & 30 \\
Early stopping improvement threshold & 1\% \\ 
Filter size & 5 \\
Dropout rate & 0.2 \\
Total number of parameters & 63M \\
  \end{tabular}
\end{ruledtabular}
\end{table}

\begin{table}
\caption{\label{tab:unetparam} U-Net Hyperparameters}
\begin{ruledtabular}
  \begin{tabular}{lc}
Initial learning rate  & $10^{-3}$ \\
Max epochs & 200 \\
Batch size & 32 \\
Train set size & 3488 \\
Test set size & 832 \\
Optimizer & AdamW \\
Learning rate scheduler & CosineAnnealingLR w/ warmup \\
Learning rate scheduler warmup epochs & 5 \\
Weight decay & $10^{-4}$ \\
Gradient clipping & 0.50 \\
Early stopping patience & 30 \\
Early stopping improvement threshold & 1\% \\ 
Filter size & 5 \\
Dropout rate & 0.2 \\
Total number of parameters & 58M \\
  \end{tabular}
  \end{ruledtabular}
\end{table}

\begin{table}
\caption{\label{tab:fnoparam}FNO Hyperparameters}
\begin{ruledtabular}
\begin{tabular}{lc}
Initial learning rate  & $10^{-3}$ \\
Max epochs & 200 \\
Batch size & 32 \\
Train set size & 3488 \\
Test set size & 832 \\
Optimizer & AdamW \\
Learning rate scheduler & CosineAnnealingLR w/ warmup \\
Learning rate scheduler warmup epochs & 5 \\
Weight decay & $10^{-4}$ \\
Gradient clipping & 0.50 \\
Early stopping patience & 30 \\
Early stopping improvement threshold & 1\% \\ 
Fourier layer width & 80 \\
Fourier layer modes & 24 \\
Total number of parameters & 59M \\
\end{tabular}
\end{ruledtabular}
\end{table}

\begin{figure}
  \centerline{\includegraphics[width=\textwidth]{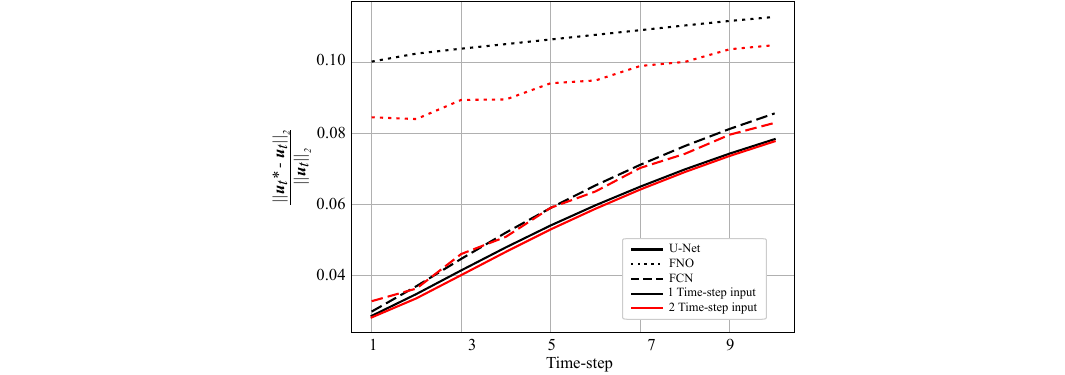}}
  \caption{Relationship between error and time-step for 
  \rev{deep learning} models with and without a time-delay embedding at $Re=590$.}
\label{fig:multistep_error}
\end{figure}

\begin{figure}
  \centerline{\includegraphics[width=\textwidth]{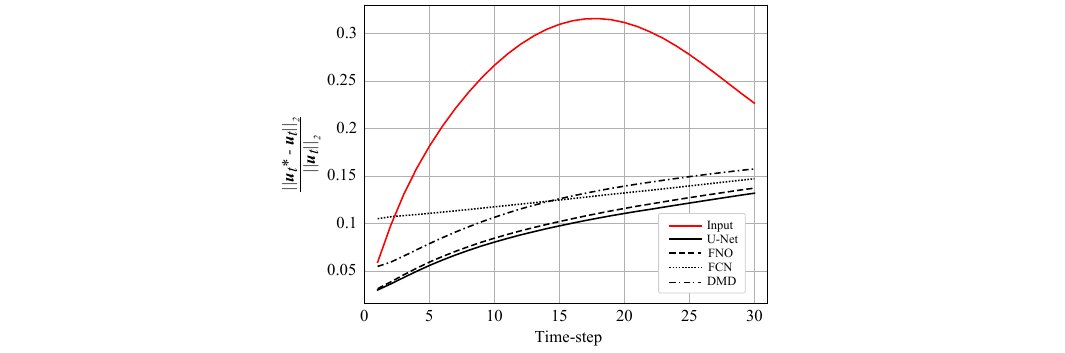}}
  \caption{\rev{Mean error per time-step for various models at $Re = 590$ compared with persistence baseline. The thirty time-step horizon shown is approximately equal to one full vortex shedding cycle, resulting in the error decreasing in the second half of the forecast as the measured flow moves back into phase with the initial observation.}}
\label{fig:error_vs_time_pers}
\end{figure}

\begin{figure}
  \centerline{\includegraphics[width=\textwidth]{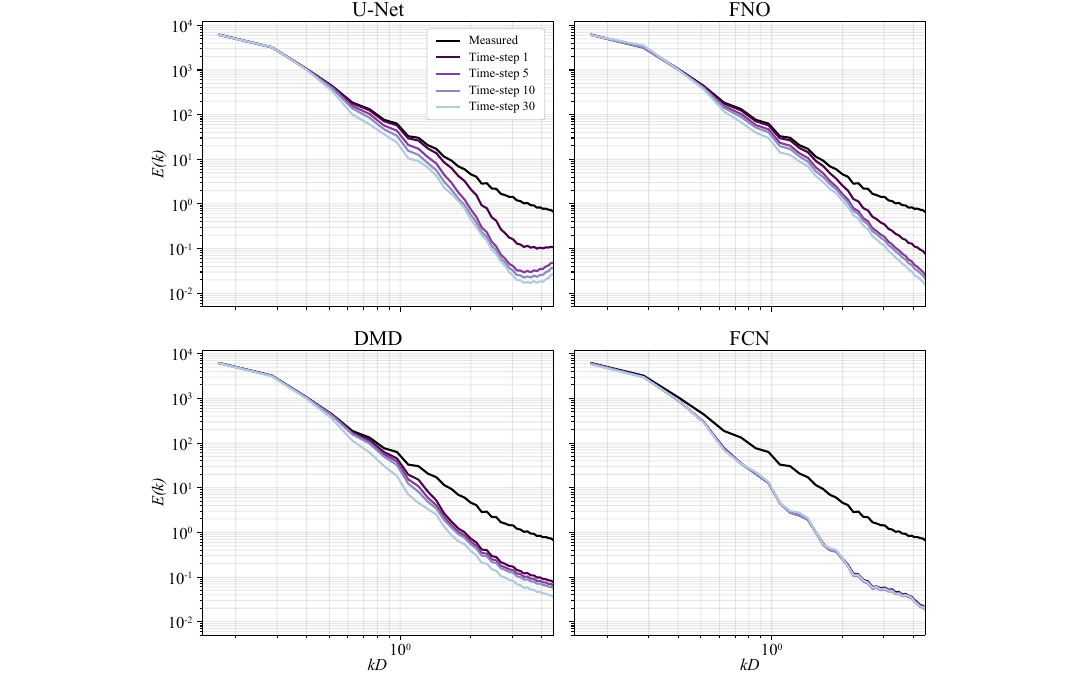}}
  \caption{\rev{Radially averaged spatial energy spectra of measured and forecast velocity fields at $Re=590$.}}
\label{fig:energy_spectra}
\end{figure}

\newpage
\cleardoublepage
\bibliography{arxiv_temp}

\end{document}